\documentclass[pra,aps,superscriptaddress,balancelastpage,onecolumn]{revtex4-2}
\usepackage{amsmath}
\usepackage{graphicx}
\usepackage{xcolor}
\usepackage{amsfonts}
\usepackage{amssymb}
\usepackage{bbold}
\providecommand{\U}[1]{\protect\rule{.1in}{.1in}}
\providecommand{\U}[1]{\protect\rule{.1in}{.1in}}

\begin{document}

\title{Distant entanglement via photon hopping in a coupled magnomechanical system}
\author{Amjad Sohail}
\affiliation{Department of Physics, Government College University, Allama Iqbal Road, Faisalabad 38000, Pakistan}
\affiliation{Electrical and Computer Engineering Department, Abu Dhabi University, Abu Dhabi 59911, United Arab Emirates}
\author{Jia-Xin Peng}
\affiliation{State Key Laboratory of Precision Spectroscopy, Quantum Institute for Light and Atoms, Department of Physics, East China Normal University, Shanghai 200062, China}
\author{Abdelkader Hidki}
\affiliation{LPTHE, Department of Physics, Faculty of Sciences, Ibn Zohr University, Agadir, Morocco}
\author{S. K. Singh}
\affiliation{Graphene and Advanced 2D Materials Research Group (GAMRG), School of Engineering and Technology, Sunway University, No. 5,
Jalan Universiti, Bandar Sunway, 47500 Petaling Jaya, Selangor, Malaysia}

\begin{abstract}
We theoretically propose a scheme to generate distant bipartite entanglement between various subsystems in coupled magnomechanical systems where both the microwave cavities are coupled through single photon hopping parameter. Each cavity also contains a magnon mode and phonon mode and this gives five excitation modes in our model Hamiltonian which are cavity-1 photons, cavity-2 photons, magnon, and phonon modes in both YIG spheres. We found that significant bipartite entanglement exists between indirectly coupled subsystems in coupled microwave cavities for an appropriate set of parameters regime. Moreover, we also obtain suitable cavity and magnon detuning parameters for a significant distant bipartite entanglement in different bipartitions. In addition, it can be seen that a single photon hopping parameter significantly affects both the degree as well as the transfer of quantum entanglement between various bipartitions. Hence, our present study related to coupled microwave cavity magnomechanical configuration will open new perspectives in coherent control of various quantum correlations including quantum state transfer among macroscopic quantum systems
\end{abstract}

\maketitle


\section{Introduction}
Quantum entanglement is one of the most fascinating phenomena in quantum mechanics which is also unique property of quantum manybody  systems \cite{schwabl2007quantum,ballentine2014quantum}. In the beginning era of quantum technology, seminal theoretical and experimental investigations mainly explored  only microscopic systems, such as atoms, ions, etc to obtain  quantum entanglement \cite{horodecki2009quantum}. However, there is no clear physical law that states that quantum entanglement can only occur in microscopic systems. In 2007,  Vitali et al. first  proposed the entanglement between a single cavity mode and a vibrating mirror, which is the beginning of macroscopic entanglement research \cite{vitali2007optomechanical}. Subsequently, the study of macroscopic quantum entanglement phenomena based on optical mechanical systems received widespread attention, including the entanglement of two vibrating mirrors \cite{mancini2002entangling,yang2015generation,li2017enhanced,hartmann2008steady,liao2014entangling}, entanglement of multiple cavity modes coupled to vibrating objects \cite{xiong2005correlated,kiffner2007two,qamar2009entangled}, entanglement in Laguerre-Gaussian cavity system \cite{ChenZhen,MBBB,JiaXin,singh2021entanglement,CHJ}.

Recently, ferrimagnetic materials have provided a powerful platform for studying the essence of magnetic systems \cite{rameshti2022cavity,rao2021interferometric,zhang2021parity,shen2021long}. Yttrium iron garnet (YIG) crystal is one of the most representative materials in low damping magnetic materials owing to  its extremely high spin density and excellent integration performance \cite{tabuchi2014hybridizing,zhang2014strongly}. Particularly, the Kittel mode  in the YIG sphere and the microwave cavity photons can be coupled to achieve the vacuum Rabi splitting and  cavity-magnon polaritons \cite{kittel1948theory,li2023squeezing,li2018magnon}. This induced the birth of magnon cavity  QED, which provides a  promising platform for the study of strong interactions between light and matter. Naturally, many interesting quantum phenomena have been studied based on cavity magnetic systems, such as magnoninduced transparency \cite{ullah2020tunable,sohail2023controllable,liu2023generation,singh2023enhanced}, coherent feedback \cite{amazioug2023enhancement},  magnon dark modes \cite{zhang2015magnon}, bistability \cite{shen2022mechanical,wang2018bistability}, the magnon Kerr effect \cite{kong2019magnon,zhang2023detection,xiong2022strong,wang2016magnon,zhang2019theory}, microwave-optical conversion \cite{hisatomi}, magnon blocking \cite{yan2020magnon,liu2019magnon}, and so on.
 
It is worth mentioning that, Li Jie first studied magnon-photon-phonon entanglement based on cavity magnetic system in 2018, which opens a new subfield within the field of quantum entanglement \cite{li2018magnon}. Subsequently, them presentd a scheme to entangle two microwave fields by using the nonlinear magnetostrictive interaction \cite{yu2020magnetostrictively,hidki2023evolution}. In addition, the macroscopic entanglement between two YIG spheres has also been studied \cite{li2019entangling}. Considering that the the magnon Kerr effect  may be boost the quantum effect, this stimulates the research on enhancing the entanglement between two magnon modes by using the Kerr effect \cite{zhang2019quantum,yang2020entanglement}.  Moreover, the photon-magnon entanglement is improved by using parametric amplifier \cite{hussain2022entanglement,hidki2023enhanced} and squeezing effect \cite{sohail2023enhanced,hidki2023transfer}.  More interestingly, the remote magnon entanglement between two massive ferrimagnetic spheres \cite{wu2021remote,sun2021remote} and 
robust optical entanglement are also implemented in cavity optomagnonics system \cite{xie2023generation}. 
Besides the entanglement between magnon and ordinary cavity mode, the entanglement between magnon and Laguerre-Gaussian cavity mode is also been studied theoretically \cite{CHJ}. 
Furthermore,  distant entanglement via photon hoping between different modes has been of great interest for storing/sharing quantum information. Recently, Chen et. al. studied the perfect transferring of enatnaglement and quntum steering between different modes in coupled cavity magnomechanical system \cite{chen2021perfect}. In addition, Dilawaiz et. al. investigate the entanglement between a YIG sphere and an atomic ensamble via photon hoping in coupled microwave cavities by \cite{dilawaiz2022entangled}. 
Therefore, researchers pay more attention to investigate the quantum corelation via photon hoping among different/distant bipartitions, 
Motivated by these developments, we consider coupled magnomechanical system to investigate weather we can generate distant enatnglement between different bi-partitions. Therefore, we emphsis on the underlying physical understanding of the generation of distant entanglements via photon hoping. Furthermore, such a well-designed coupled magnomechanical system can be utilized to create and transfer continuous variable entanglement between different distant bosonic modes.

\section{The Model}
The magnomechanical system under consideration consists of two MW cavities connected through single photon hoping factor $\Gamma$. Each cavity contain a magnon mode $m$ and a phonon mode $b$ as shown in Fig. 1. The magnons are considered to be quasiparticles which are incorporated by a collective excitation of a large number of spins inside a ferrimagnet, e.g., a YIG sphere \cite{1}. The magnetic dipole interaction enables the coupling between the magnon and the MW field. The orientation of YIG sphere inside each cavity field is in the region of the maximum magnetic field (See Fig. 1). 
At the YIG sphere site, the magnetic field of the cavity mode is along the x axis while the drive magnetic field is along the y direction). Furthermore, the bias magnetic field is set in the z direction.
In addition, the magnon and phonon modes are coupled to each other via magnetostrictive force, which yields the magnon-phonon
coupling \cite{mag3,Kittel}. The magnetostrictive interaction
depends on the resonance frequencies of the magnon and phonon modes \cite{Zha}. In the current study, we assumed the frequency magnon to be much larger than mechanical frequency, which helps to set up the strong
dispersive magnon-phonon interaction \cite{1,Ballestero}. The
Hamiltonian of the magnomechanical system can be written as
\begin{eqnarray}
H/\hbar &=&H_{0}+H_{int}+H_{d},
\end{eqnarray}%
where
\begin{eqnarray}
H_{0} &=&\sum_{k=1}^{2}\left[\omega _{k}c_{k}c_{k}^{\dag }+\omega _{m_{k}}m_{k}^{\dag }m_{k}+%
\frac{\omega _{b_{k}}}{2}\left( q_{k}^{2}+p_{k}^{2}\right)\right], \\
H_{int} &=&\sum_{k=1}^{2}\left[[g_{mb}m^{\dag }m_{k}q_{k} +g_{k}\left( c_{k}m^{\dag
}+c_{k}^{\dag }m\right)\right] +\Gamma(c_{1}c_{2}^{\dag }+c_{1}^{\dag }c_{2}), \\
H_{d}&=&i\Omega \sum_{k=1}^{2}\left[] m_{k}^{\dag }e^{-i\omega _{0}t}-m_{k}e^{i\omega _{0}t}\right],
\label{HM}
\end{eqnarray}%
where $c_{k}\left( c_{k}^{\dag }\right) $ and $m_{k}\left( m_{k}^{\dag }\right) $  
are the annihilation (creation) operator of the the $k$ cavity and magnon mode, respectively. Furthermore, $q_{k}$ and $p_{k}$ are the position and momentum quadratures of the respective mechanical mode of the magnon. In addition $\omega _{k}$, $\omega _{b}$ and $%
\omega _{m}$ are the resonance frequencies of the cavity mode $k$,
mechanical mode and the magnon mode. The magnon frequency $\omega _{m}$ can
be flexibly adjusted by the bias magnetic field $B$ via $\omega _{m}=\gamma
_{0}B$. Here $\gamma _{0}$ is the gyromagnetic ratio. The optomagnonical
coupling is theoretically given by
\begin{eqnarray}
\Gamma_{k}&=&\mathcal{V}\frac{c}{n_{r}}\sqrt{\frac{2}{\rho_{spin}V_{YS}}},
\end{eqnarray}
where $\mathcal{V}$, $n_{r}$, $\rho_{spin}$ and $V_{YS}=\frac{4\pi r^{3}}{3}$
are, respectively, the YIG sphere's Verdet constant, the refractive index,
the spin density and the volume of the YIG sphere \cite{AORH}. We considered
strong coupling regime i.e., the coupling between the cavity mode $k$ with
magnon mode $\Gamma_{k} $ can be larger than the decay rate of the magnon
and the cavity modes, $\Gamma_{k}>\kappa_{k}$, $\kappa_{m}$ \cite%
{kitt2,Tabuchi,kitt3,Goryachev}. Here, $g_{mb}$ denotes single-magnon
magnomechanical coupling rate which is considered to be very small but can
be enhanced by directly driving the YIG sphere with a MW source. The Rabi
frequency $\Omega =\left( \sqrt{5}/4\right) \gamma _{0}\sqrt{N_{spin}}B_{0}$
\cite{Simon,Holstein} represents the coupling strength of the drive
field with frequency $\omega _{0}$), amplitude $B_{0}=3.9\times 10^{-9}$T,
where $\gamma_{0}=28$GHz/T and the total number of spins $N_{spin}=\rho
V_{YS}$ with the spin density of the YIG $\rho_{spin} =4.22\times
10^{27}m^{-3}$. In addition, it is noteworthy to mention here that
collective motion of the spins are truncated to form bosonic operators $m$
and $m^{\dagger}$ via the Holstein-Primakoff transformation and further, the
Rabi frequency $\Omega $ is derived under the basic assumption of the
low-lying excitations $2Ns\gg\langle m^{\dagger}m \rangle$, where $s=\frac{5%
}{2}$ is the spin number of the ground state Fe$^{3+}$ ion in YIG.
\begin{figure}[tbp]
\begin{center}
\includegraphics[width=0.8\columnwidth,height=4in]{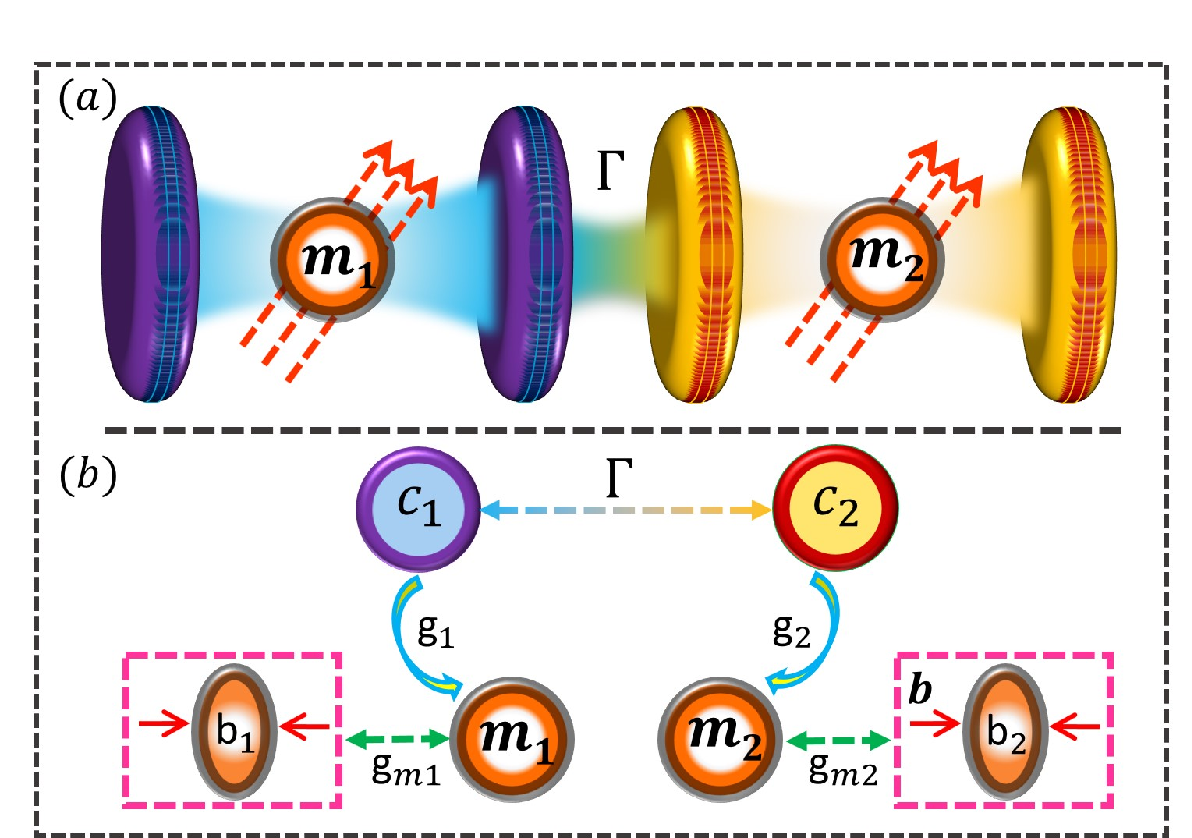}
\end{center}
\caption{(Color Online) (a) Schematic diagram of the coupled cavity magnomechanical system in which each cavity mode contain a magnon mode in a YIG sphere couples that interact with the microwave cavity modes via magnetic dipole interaction and with phonon mode via magnetostrictive interaction. The magnetic field of the each microwave cavity modes is set to be along the x-direction, while the drive magnetic field (bias magnetic field) is considered along y-direction (z-direction).
(b) The linear coupling diagram of each cavity magnomechanical system is shown. The two cavity modes are coupled via photon hoping $\Gamma$, while a cavity mode photon $c_{1}$ ($c_{2}$) is coupled to the magnon mode $m_{1}$ ($m_{2}$), with
coupling strength $g_{1}$ ($g_{2}$), which then coupled to a phonon mode $b_{1}$ ($b_{2}$) to with magnomechanical coupling strength $g_{m1}$ ($g_{m2}$).}
\end{figure}

In the rotating wave approximation at the drive frequency $\omega _{0}$, he
Hamiltonian of the system can be written as%
\begin{eqnarray}
H/\hbar &=&\sum_{k=1}^{2}\left[\Delta _{k}c_{k}c_{k}^{\dag }+\Delta
_{m_{k}}m_{k}^{\dag }m_{k}+\frac{\omega _{b_{k}}}{2}\left(
q_{k}^{2}+p_{k}^{2}\right)+g_{mk}m_{k}^{\dag}m_{k}q_{k} +g_{k}\left(
c_{k}m_{k}^{\dag } +c_{k}^{\dag }m_{k}\right)+i\Omega\left( m_{k}^{\dag
}-m_{k}\right)\right] \\
&&+\Gamma(c_{1}c_{2}^{\dag }+c_{1}^{\dag }c_{2}),  \notag
\end{eqnarray}
where $\Delta _{k}=\omega _{k}-\omega _{0k}$ ($k=1,2$) and $%
\Delta_{m_{k}}=\omega _{m}-\omega _{0k}$.

\section{Quantum dynamics and entanglement of the magnomechanical system}

We now start to obtain the equations for the dynamics of this
magnomechanical system. By incorporating the effect of noises and
dissipations, the following set of quantum Langevin equations for the
magnomechanical system can be obtained:
\begin{eqnarray}
\dot{q}_{k} &=&\omega _{b_{k}}p_{k},  \label{LG} \\
\dot{p}_{k} &=&-\omega _{b}q_{k}-\gamma
_{b}p_{k}-g_{mk}m^{\dag}_{k}m_{k}+\xi_{k} , \\
\dot{c}_{k} &=&-\left( i\Delta _{k}+\kappa_{k}\right)
c_{k}-ig_{k}m_{k}+\Gamma c_{j}+ \sqrt{2\kappa_{a}}c_{k}^{in}, (j\neq k) \\
\dot{m}_{k} &=&-\left( i\Delta
_{m_{k}}+\kappa_{m}\right)m_{k}-ig_{k}c_{k}-ig_{mk}m_{k}q_{k}+\Omega_{k} +%
\sqrt{2\kappa_{m}}m^{in},  \label{LEE}
\end{eqnarray}%
where $k_{k}$($k_{m}$) is the decay rate of the $k^{th}$ cavity mode (magnon
mode) while $\gamma _{b}$ denotes the mechanical damping rate. $\xi $, $%
m^{in}$ and $c_{k}^{in}$ are input noise operators for the mechanical,
magnon and cavity modes respectively. These noise operators are
characterized by the following correlation functions \cite{Gardiner}:
\begin{eqnarray}
\left\langle \xi (t)\xi (t^{\prime })\right\rangle +\left\langle \xi
(t^{\prime })\xi (t)\right\rangle /2&=& \gamma_{b} \lbrack 2n_{b}(\omega
_{b})+1]\delta (t-t^{\prime }), \\
\left\langle c_{k}^{in}(t)c_{k}^{in\dag }(t^{\prime })\right\rangle
&=&[n_{k}(\omega _{k})+1]\delta (t-t^{\prime }), \\
\left\langle c_{k}^{in\dag }(t)c_{k}^{in}(t^{\prime })\right\rangle
&=&n_{k}(\omega _{k})\delta (t-t^{\prime }), \\
\left\langle m^{in\dag }(t)m^{in}(t^{\prime })\right\rangle
&=&n_{m}(\omega_{m})\delta (t-t^{\prime }), \\
\left\langle m^{in}(t)m^{in\dag }(t^{\prime })\right\rangle &=&[n_{m}(\omega
_{m})+1]\delta (t-t^{\prime }),  \label{LF}
\end{eqnarray}
The equilibrium mean thermal photon, magnon, and phonon numbers are $%
n_{f}(\omega _{f})=[\exp (\frac{\hbar\omega_{f}}{k_{b}T})-1]^{-1}$ $%
[f=k(k=1,2),m,b]$, where $T$ is the environmental temperature and $k_{b}$
the Boltzmann constant.

If the magnon mode is strongly driven, then we must have $\left\vert
\left\langle m\right\rangle \right\vert \gg 1$. In addition, the two MW
cavity fields show large amplitudes due to the cavity-magnon beam splitter
interactions. This permits us to linearize the above quantum Langevin
equations by writing any operator as a sum of average value plus its
fluctuation i.e., $o=\left\langle o\right\rangle +\delta o$, $(o=p,q,c_{k},m)
$ and substitute it into Eq.(\ref{LG}-\ref{LEE}). The average values of the
dynamical operators are obtained as
\begin{eqnarray}
\left\langle p _{k}\right\rangle  &=&0, \\
\left\langle q _{k}\right\rangle  &=&\frac{-g_{mk}}{\omega _{b}}\left\vert
\left\langle m _{k}\right\rangle \right\vert ^{2}, \\
\left\langle m_{j}\right\rangle  &=&\frac{\Omega _{j}-ig _{k}\left\langle
c _{k}\right\rangle }{i\Delta _{m _{k}}+\kappa _{m _{k}}}, \\
\left\langle c_{1}\right\rangle  &=&\frac{ig_{1}\Omega _{1}\alpha
_{2}-\Gamma g_{2}\Omega _{2}(\kappa _{m_{1}}+i\Delta _{m_{1}})}{\alpha
_{1}\alpha _{2}+\Gamma ^{2}\left( i\Delta _{m_{1}}+\kappa _{m_{1}}\right)
\left( i\Delta _{m_{2}}+\kappa _{m_{2}}\right) }, \\
\left\langle c_{2}\right\rangle  &=&\frac{ig_{2}\Omega _{2}\alpha
_{1}-\Gamma g_{1}\Omega _{1}(\kappa _{m_{2}}+i\Delta _{m_{2}})}{\alpha
_{1}\alpha _{2}+\Gamma ^{2}\left( i\Delta _{m_{2}}+\kappa _{m_{2}}\right)
\left( i\Delta _{m_{1}}+\kappa _{m_{1}}\right) },
\end{eqnarray}%
where $\alpha _{j}=(i\Delta _{i}+\kappa _{i})(i\Delta _{m_{i}}+\kappa
_{m_{i}})+g_{i}^{2}$ , $\Delta _{m}=\Delta _{m_{0}}+g_{mb}\left\langle
q\right\rangle $ is the effective magnon mode detuning which includes the
slight shift of frequency due to the magnomechanical interaction.

Now, we introduce the quadrature for the linearised quantum Langevin equations describing
fluctuations are: $\delta x=\frac{1}{\sqrt{2}}(\delta m-\delta m^{\dag })$, $\delta y=\frac{1}{%
\sqrt{2}i}(\delta m-\delta m^{\dag })$, $\delta X_{k}=\frac{1}{\sqrt{2}}%
(\delta c_{k}-\delta c_{k}^{\dag })$, $\delta Y_{k}=\frac{1}{\sqrt{2}i}%
(\delta c_{k}-\delta c_{k}^{\dag })$ can be written in concise form as%
\begin{equation}
\dot{\mathcal{F}}(t)=\mathcal{M}\mathcal{F}(t)+\mathcal{N}(t),
\end{equation}%
where $\mathcal{F}(t)$ and $\mathcal{N}(t)$ are, respectively, the quantum
the fluctuation and input noise vectors and are given by:%
\begin{eqnarray*}
\mathcal{F}(t) &=&[\delta C_{XY}(t),\delta M_{xy}(t),\delta Q_{qp}(t)]^{T},
\\
\mathcal{N}(t) &=&[\mathcal{N}_{XY},\mathcal{N}_{xy},\mathcal{N}_{qp}]
\end{eqnarray*}%
where%
\begin{eqnarray*}
\delta C_{XY}(t) &=&\delta X_{1}(t),\delta Y_{1}(t),\delta X_{2}(t),\delta
Y_{2}(t) \\
\delta M_{xy}(t) &=&\delta x_{1}(t),\delta y_{1}(t),\delta x_{2}(t),\delta
y_{2}(t) \\
\delta Q_{qp}(t) &=&\delta q_{1}(t),\delta p_{1}(t),\delta q_{2}(t),\delta
p_{2}(t)
\end{eqnarray*}%
\begin{eqnarray*}
\mathcal{N}_{XY} &=&\sqrt{2k_{1}}X_{1}^{in}(t),\sqrt{2k_{1}}Y_{1}^{in}(t),%
\sqrt{2k_{2}}X_{2}^{in}(t),\sqrt{2k_{2}}Y_{2}^{in}(t) \\
\mathcal{N}_{xy} &=&\sqrt{2k_{m}}x_{1}^{in}(t),\sqrt{2k_{m}}y_{1}^{in}(t),%
\sqrt{2k_{m}}x_{2}^{in}(t),\sqrt{2k_{m}}y_{2}^{in}(t) \\
\mathcal{N}_{qp} &=&0,\xi _{1}(t),0,\xi _{2}(t)
\end{eqnarray*}%
Furthermore, the drift matrix $\mathcal{M}$ can be written as
\begin{equation}
\mathcal{M}=\left[
\begin{array}{cccccccccccc}
-\kappa _{1} & \Delta _{1} & 0 & \Gamma  & 0 & g_{1} & 0 & 0 & 0 & 0 & 0 & 0
\\
-\Delta _{1} & -\kappa _{1} & -\Gamma  & 0 & -g_{1} & 0 & 0 & 0 & 0 & 0 & 0
& 0 \\
0 & \Gamma  & -\kappa _{2} & \Delta _{2} & 0 & 0 & 0 & g_{2} & 0 & 0 & 0 & 0
\\
-\Gamma  & 0 & -\Delta _{2} & -\kappa _{2} & 0 & 0 & -g_{2} & 0 & 0 & 0 & 0
& 0 \\
0 & g_{1} & 0 & 0 & -\kappa _{m_{1}} & \Delta _{m_{1}} & 0 & 0 & -G_{1} & 0
& 0 & 0 \\
-g_{1} & 0 & 0 & 0 & -\Delta _{m_{1}} & -\kappa _{m_{1}} & 0 & 0 & 0 & 0 & 0
& 0 \\
0 & 0 & 0 & g_{2} & 0 & 0 & -\kappa _{m_{2}} & \Delta _{m_{2}} & 0 & 0 &
-G_{2} & 0 \\
0 & 0 & -g_{2} & 0 & 0 & 0 & -\Delta _{m_{2}} & -\kappa _{b_{2}} & 0 & 0 & 0
& 0 \\
0 & 0 & 0 & 0 & 0 & 0 & 0 & 0 & 0 & \omega _{b_{1}} & 0 & 0 \\
0 & 0 & 0 & 0 & 0 & G_{1} & 0 & 0 & -\omega _{b_{1}} & -\gamma _{b_{1}} & 0
& 0 \\
0 & 0 & 0 & 0 & 0 & 0 & 0 & 0 & 0 & 0 & 0 & \omega _{b_{2}} \\
0 & 0 & 0 & 0 & 0 & 0 & 0 & 0 & G_{2} & 0 & 0 & \gamma _{b_{2}}%
\end{array}%
\right] ,  \label{DfM}
\end{equation}%
%
%
where $G_{mb}=i\sqrt{2}g_{mb}\left\langle m\right\rangle $ is the effective
magnomechanical coupling rate. By using Eq. (\ref{HM}), one can notice that
effective magnomechanical coupling rate can be increased by applying strong
magnon drive.

Next, we discuss the quantum correlation of bipartite subsystems with a
special emphasis on the entanglement of two indirectly coupled modes and the
two MW fields in the steady-state. The stability of the proposed system is
the first prerequisite for the effectiveness of our scheme. According to
Routh-Hurwitz criterion \cite{RHCr}, the system is stable only if the real
part of the all the eigenvalues of the drift matrix $\mathcal{M}$ are
negative. Hence, we start our analysis by determining eigenvalues of the
drift matrix $\mathcal{M}$ (i.e., $|\mathcal{M}-\lambda_{\mathcal{M}}
\mathbb{1}|=0$) and make sure the stability condition are all satisfied in
the following section (see Appendix A). The magnomechanical system presented
here is characterized by $8\times 8$ covariance matrix V with its entries
\begin{equation}
V_{ij}(t)=\frac{1}{2}\left\langle F_{i}(t)F_{j}(t^{\prime
})+F_{j}(t^{\prime})F_{i}(t)\right\rangle,  \label{CoM}
\end{equation}
The covariance matrix of the magnomechanical system can be obtained from the
steady state Lyapunov equation \cite{Parks,SA}
\begin{equation}
\mathcal{M}V+V\mathcal{M}^{T}=-\mathcal{D},  \label{f}
\end{equation}%
where $\mathcal{D}=$ diag$[0,\gamma _{b}\left( 2n_{b}+1\right) ,\kappa
_{m}\left( 2n_{m}+1\right) ,\kappa _{m}(2n_{m}+1)$ $\kappa _{1}\left(
2n_{1}+1\right)$ ,$\kappa _{1}\left( 2n_{1}+1\right) $, $\kappa _{2}\left(
2n_{2}+1\right) ,\kappa _{2}\left( 2n_{2}+1\right) ]$, is a diagonal matrix
which is called diffusion matrix and characterizes the noise correlations.
The Lyapunov Eq. (\ref{f}) as a linear equation for $V$ can be easily
solved. Using the Simon condition for Gaussian states, we calculate the
entanglement of the steady state \cite{Gonzalez,Vidal,Plenio,Adesso,SA}.
\begin{equation}
E_{N}=\max [0,-\ln 2 \eta ^{-}],
\end{equation}%
where $\eta ^{-}=$min eig$|\bigoplus^{2}_{j=1}(-\sigma_{y})\widetilde{%
\mathcal{V}_{4}}| $ is the minimum symplectic eigenvalue of covariance
matrix and is $\widetilde{\mathcal{V}_{4}}=\varrho_{1|2}\mathcal{V}%
_{in}\varrho_{1|2}$, where $\mathcal{V}_{in}$ is a $4\times4$ matrix of any
two subsystems which can easily be obtain by neglecting the uninteresting
rows and columns in $\mathcal{V}_{4}$. $\varrho_{1|2}=\sigma_{z}\bigoplus 1$%
=diag$(1,-1,1,1)$ is the matrix which characterizes the partial
transposition at the level of covariance matrices. Here, $\sigma_{y}$ and $%
\sigma_{z}$ are the pauli spin matrices. Furthermore, a nonzero logarithmic
negativity i.e., $E_{N}>0$ defines the presence of bipartite entanglement in
our cavity magnomechanical system.

\section{Results and Discussion}
\begin{table}
\centering
\begin{tabular*}{1\textwidth}{@{\extracolsep{\fill}}|l c c ||l c c|}
\hline
Parameters & Symbol & Value & Parameters & Symbol & Value \tabularnewline
\hline\hline
Phonon frequency  & $\omega_{b}$ & $2\pi \times $10 MHz & Cavity frequency  & $\omega_{1}=\omega_{2}=\omega_{a}$ & $2\pi \times $ 10 GHz
\tabularnewline
\hline
Cavity decay rates  & $\kappa_{1}=\kappa_{2}=\kappa$ & $2\pi \times $ 1 MHz & Magnon decay rate  & $\kappa_{m}$ & $2\pi  \times $1 MHz
 \tabularnewline
\hline
Mechanical damping rate & $\gamma_{b}$ & $2\pi \times $100 Hz & Magnon-Microwave couplings & $\Gamma_{1}=\Gamma_{2}=\Gamma$ & $2\pi \times $3.2 MHz\tabularnewline
\hline
Magnomechanical coupling  & $g_{mb}$ & $2\pi \times $0.3 Hz & Drive Magnetic Field  & B & $3.9\times10^{-5}$T \tabularnewline
\hline
YIG Sphere Diameter  & D & 250 $\mu$m & Temperature  & T & 10 mK \tabularnewline
\hline
Power  & $\wp=\frac{B^{2}\pi r^{2} c}{2\mu^{2}}$ & 9.8 mW & Spin density & $\rho$ & $4.22\times 10^{27}m^{-3}$ \tabularnewline
\hline
\end{tabular*}
\caption{Parameters used in recent experiments for the mechanical resonators} 
\label{table:pvalue}
\end{table}
\begin{table}
\centering
\begin{tabular*}{0.8\textwidth}{@{\extracolsep{\fill}}|l c ||c c |}
\hline
Bipartite Subsystem & Entanglement Symbol & Bipartite Subsystem & Entanglement Symbol \tabularnewline
\hline
Cavity 1-Cavity 2  & $E_{c-c}^{N}$ & &\tabularnewline
\hline
Cavity 1-magnon 2  & $E^{N}_{c_{1}-m_{2}}$ & Cavity 2-magnon 1 & $E^{N}_{c_{2}-m_{1}}$ \tabularnewline
\hline
Cavity 1-phonon 2  & $E^{N}_{c_{1}-b_{2}}$ & Cavity 2-phonon 1  & $E^{N}_{c_{2}-b_{1}}$  \tabularnewline
\hline
\end{tabular*}
\caption{Adopted notation for the different bipartite subsystem
entanglement.} 
\label{table:pvaluee}
\end{table}
\begin{figure}[b!]
\begin{center}
\includegraphics[width=1\columnwidth,height=5in]{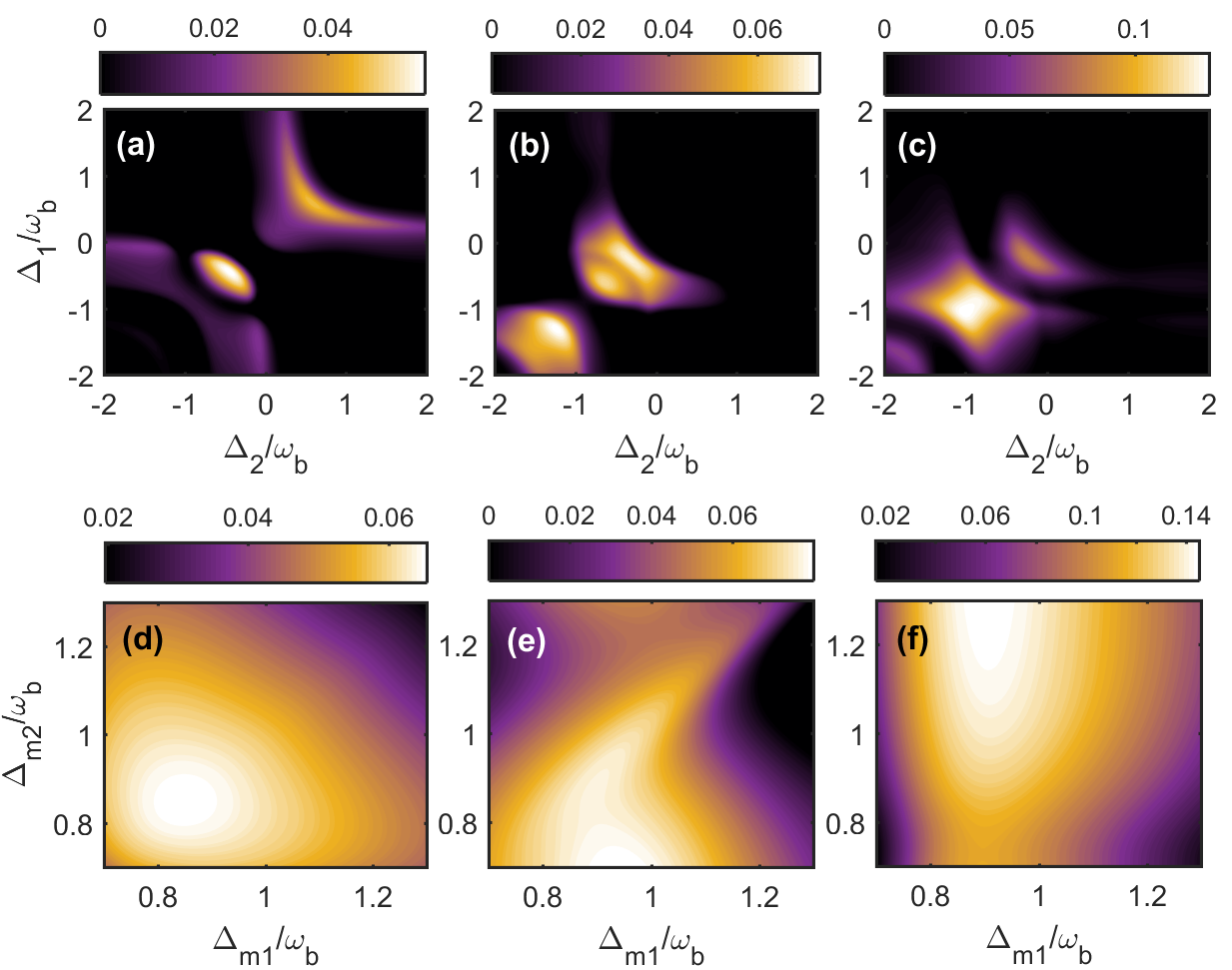}
\end{center}
\caption{(Color Online) Density plot of bipartite entanglement in (a),(d) $%
E_{c_{1}-c_{2}}^{N}$; in (b),(e) $E_{c_{1}-m_{2}}^{N}=E_{c_{2}-m_{1}}^{N}$ and in (c),(f) $E_{c_{1}-b_{2}}^{N}=E_{c_{2}-b_{1}}^{N}$ versus cavity detunings $\Delta_{1}/\protect\omega _{b}$ and $\Delta _{2}/\protect\omega _{b}$  in (a)-(c) for $\Delta_{m_{1}}=\Delta _{m_{2}}=\protect\omega _{b}$  whereas varying  both magnon detunings $\Delta_{m_{1}}/\protect\omega _{b}$ and $\Delta _{m_{2}}/\protect\omega _{b}$ in (d)-(f). We use optimum values of of   $\Delta_{1}$ and $\Delta_{2}$ for (d)-(f). The other parameters are given in Table \ref{table:pvalue}.}
\end{figure}
In this section we are going to discuss in details the results of bipartite entanglements as we have six different modes in this coupled cavity Magnomechnical system. So, we can get bipartite entanglement in any of two modes however the most significant part of our study is to investigate the bipartite  entanglement present in various spatially distant subsystems which we have summarised in Table \ref{table:pvaluee} with symbols.\\
In Fig. 2, we present five different distant bipartite entanglements as a function of dimensionless cavity detuning for first cavity  $\Delta_{1}/\protect\omega _{b}$ and second cavity  $\Delta _{2}/\protect\omega _{b}$. When both the magnon detuning is kept in resonant with blue sideband regime i.e. $\Delta_{m_{1}}=\Delta _{m_{2}}=\protect\omega _{b}$ it can be seen that bipartite entanglement between two cavity modes  $E_{c_{1}-c_{2}}^{N}$ become maximum for $\Delta _{1}=\Delta_{2}= -0.5\protect\omega _{b}$ although even if both the cavities are resonant with blue sideband regime i.e.  $\Delta _{1}=\Delta_{2}= \omega _{b}$ we have significant amount of bipartite entanglement in $E_{c_{1}-c_{2}}^{N}$ as shown in Fig. 2(a).  In Fig.2(b) we study the bipartite entanglement $E_{c_{1}-m_{2}}^{N} \textbf{(} E_{c_{2}-m_{1}}^{N}\text{)}$ which attains maximum value either when both the cavities are resonant with driving field, i.e. $\Delta _{1}=\Delta_{2}= 0$ or resonant with red sideband regime, i.e. $\Delta _{1}=\Delta_{2}= -\omega _{b}$. Moreover when both the cavities are kept in resonant with  this red sideband regime the bipartite entanglement  $E_{c_{1}-b_{2}}^{N} \textbf{(}E_{c_{2}-b_{1}}^{N}\textbf{)}$ attains its maximum value as shown in Fig. 2(c). Furthermore, it can be seen that if cavity detunigs for both the cavities are kept fixed and resonant with blue sideband regime  i.e. $\Delta _{1}=\Delta_{2}= \omega _{b}$ then all the above mentioned bipartite entanglements have significant values on gradually varying both   $\Delta_{m_{1}}/\protect\omega _{b}$ and $\Delta _{m_{2}}/\protect\omega _{b}$ from 0.8 to 1.1 as shown in Fig. 2(d)-2(f).\\
\begin{figure}[b!]
\begin{center}
\includegraphics[width=1\columnwidth,height=5in]{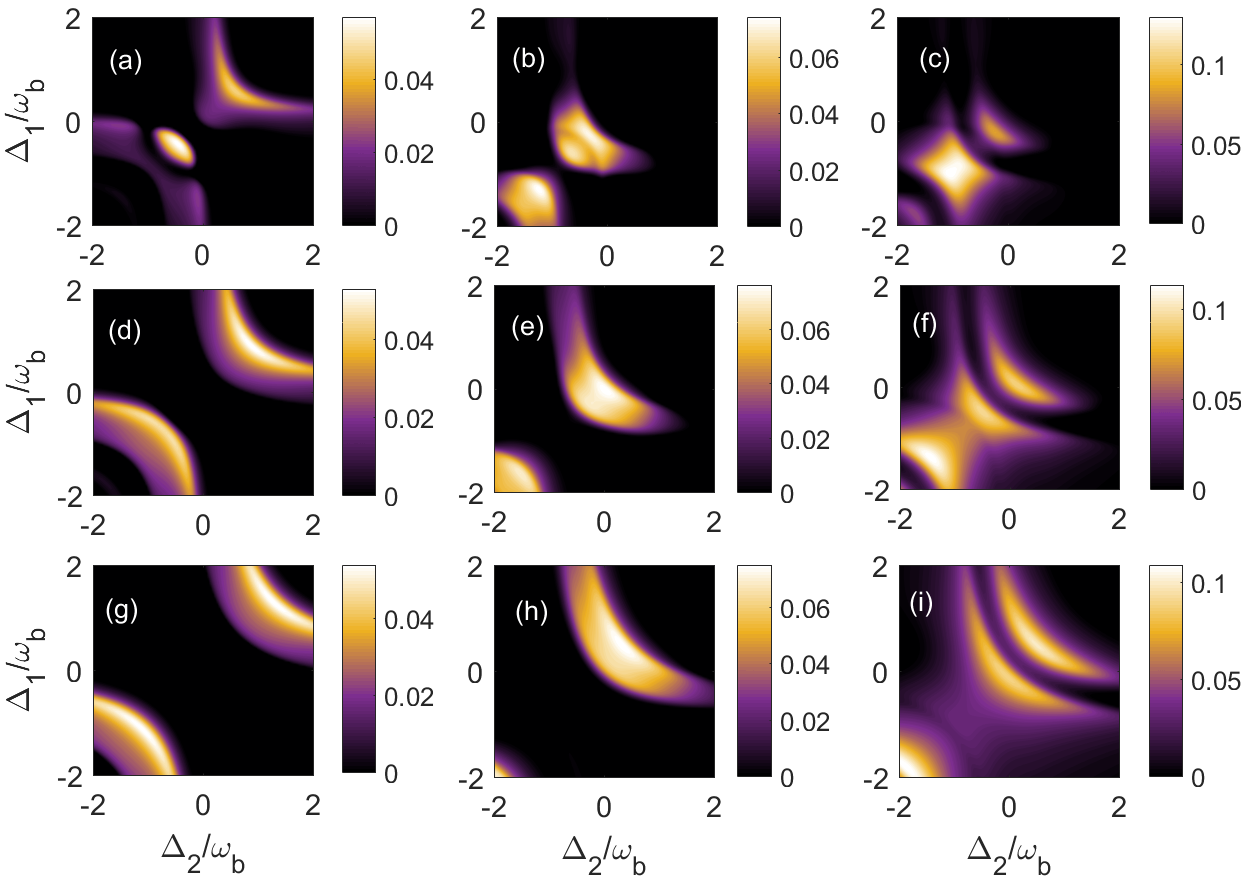}
\end{center}
\caption{(Color Online) Density plot of bipartite entanglement in (a),(d),(g) $E_{c_{1}-c_{2}}^{N}$; in (b),(e),(h) $E_{c_{1}-m_{2}}^{N}=E_{c_{2}-m_{1}}^{N}$ and in (c),(f),(i) $E_{c_{1}-b_{2}}^{N}=E_{c_{2}-b_{1}}^{N}$ versus detunings $\Delta _{1}/\protect\omega _{b}$ and $\Delta
_{2}/\protect\omega _{b}$. Here we have taken $\Gamma=0.5\omega_{b}$ for (a),(b),(c); $\Gamma=0.8\omega_{b}$ for (d),(e),(f) and $\Gamma=\omega_{b}$ for (g),(h),(i). We take both magnon detunings at  $\Delta_{m_{1}}=\Delta_{m_{2}}=\omega_{b}$ The other parameters are given in Table \ref{table:pvalue}}
\end{figure}
We plot five different distant bipartite entanglements as a function of $\Delta_{1}/\omega _{b}$ and  $\Delta _{2}/\omega _{b}$ for different photon hopping factor $\Gamma$ as well while keepin  both the magnon detuning  in resonant with blue sideband regime i.e. $\Delta_{m_{1}}=\Delta _{m_{2}}=\omega _{b}$   in Fig. 3.  For  $\Gamma = =0.5\omega _{b}$ the quantity $
E_{c_{1}-c_{2}}^{N}$ attains maximum value when both the cavity have zero detunings i.e. $\Delta_{1}=\Delta _{2} = 0$ whereas for off resonant cavities we get finite values of  $E_{c_{1}-c_{2}}^{N}$ as shown in Fig. 3(a). However the quantities  $E_{c_{1}-m_{2}}^{N}=E_{c_{2}-m_{1}}^{N}$ become maximum  for two values of cavity detunings which are  $\Delta_{1}=\Delta _{2} = 0$  and  $\Delta_{1}=\Delta _{2} = -\omega _{b}$ as shown in Fig. 3(b) whereas both the quantities $E_{c_{1}-b_{2}}^{N}=E_{c_{2}-b_{1}}^{N}$ become maximum at $\Delta_{1}=\Delta _{2} = -\omega _{b}$ as shown in Fig. 3(c). It can be seen that if we increase photon hopping factor upto  $\Gamma = = 0.8\omega _{b}$ then the bipartite entanglement in between both the cavity modes $E_{c_{1}-c_{2}}^{N}$ becomes maximum for two cases i.e. for resonant cavities  $\Delta_{1}=\Delta _{2} = 0$  and when resonant with red sideband regime $\Delta_{1}=\Delta _{2} = -\omega _{b}$ as shown in Fig. 3(d). In addition, in the density plots of the quantities  $E_{c_{1}-m_{2}}^{N}=E_{c_{2}-m_{1}}^{N}$ the panel corresponding to red sideband regime start to decrease whereas the panel corresponding to resonant cavities increases as shown in Fig. 3(e). Moreover, the quantities $E_{c_{1}-b_{2}}^{N}=E_{c_{2}-b_{1}}^{N}$ show the finite values for a broad range of cavity detunings and attain maximum value for $\Delta_{1}=\Delta _{2} = -1.5\omega _{b}$ as shown in Fig. 3(f). On further increasing the value of $\Gamma$ and keeping it at  $\Gamma = = \omega _{b}$, the quantity   $E_{c_{1}-c_{2}}^{N}$ again becomes maximum for two cases i.e. for $\Delta_{1}=\Delta _{2} =  -0.5\omega _{b}$  and  $\Delta_{1}=\Delta _{2} = -1.5\omega _{b}$ as shown in Fig. 3(g) whereas  the quantities  $E_{c_{1}-m_{2}}^{N}=E_{c_{2}-m_{1}}^{N}$ attain maximum value only when both the cavity detunings are  nearly resonant with blue sideband regime  as given in Fig. 3(h). However the  quantities $E_{c_{1}-b_{2}}^{N}=E_{c_{2}-b_{1}}^{N}$ attain maximum value only for very far off-resonant cavities $\Delta_{1}=\Delta _{2} = -2\omega _{b}$ whereas for a broad range of negative cavity detunings both these distant entanglements almost become negligible however for a positive value of $\Delta_{1}/\omega _{b}$ and $\Delta _{2}/\omega _{b}$  both the bipartite entanglements attain finite values $E_{c_{1}-b_{2}}^{N}=E_{c_{2}-b_{1}}^{N}$ as shown in Fig. 3(i).\\

\begin{figure}[tbp]
\begin{center}
\includegraphics[width=1\columnwidth,height=4.5in]{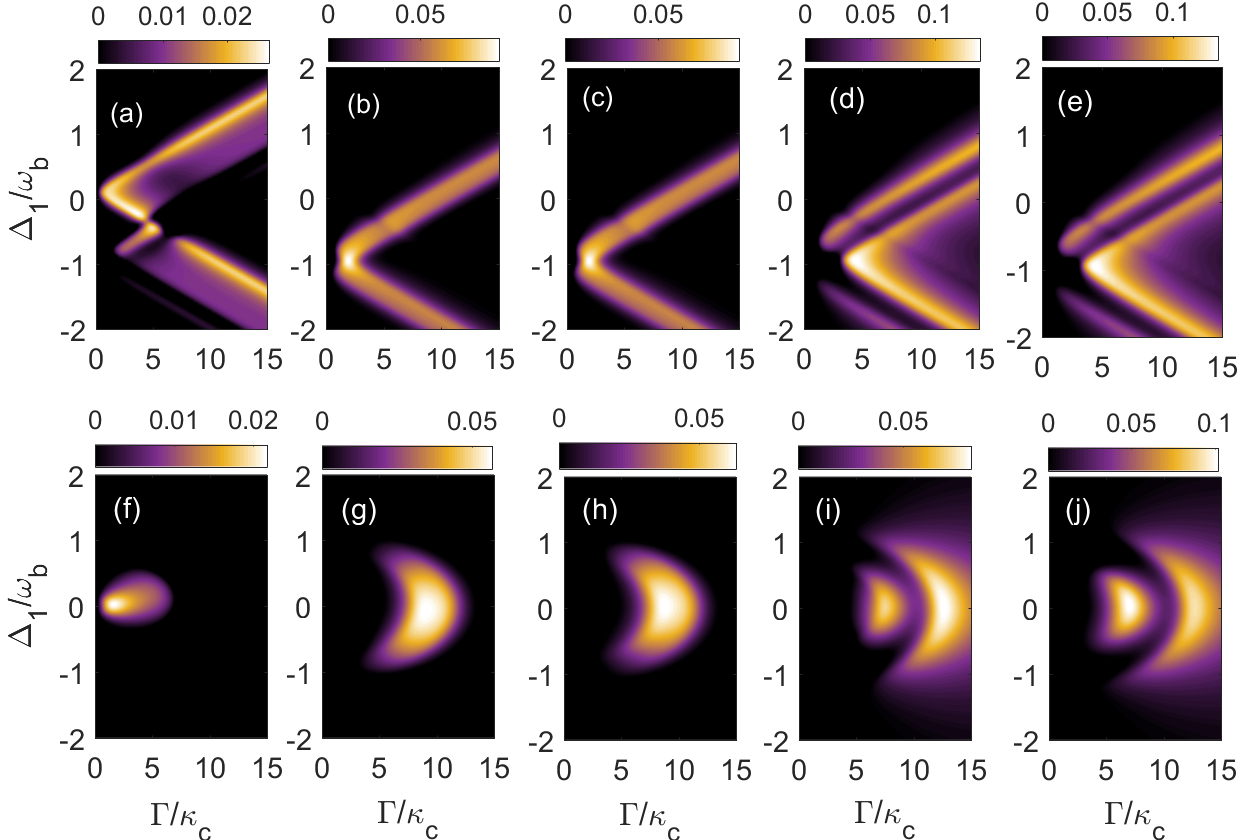}
\end{center}
\caption{(Color Online) Density plot of bipartite entanglement in (a),(f) $%
E^{N}_{c_{1}-c_{2}}$; in (b),(g) $E^{N}_{c_{1}-m_{2}}$; in (c),(h) $E^{N}_{c_{2}-m_{1}}$; in (d),(i) $E^{N}_{c_{1}-b_{2}}$ and (e),(j) $E^{N}_{c_{2}-b_{1}}$ versus $\Delta_{1}/\protect\omega_{b}$ and $\Gamma/\protect\omega_{b}$ for $\Delta_{2}=\protect\omega_{b}$ in (a)-(e) and for $\Delta_{2}=-\protect\omega_{b}$ in (f)-(j). The
other parameters are same as in Fig. 3.}
\end{figure}
We study the effects of varying photon hopping factor $\Gamma/\kappa_{c}$  and normalised cavity detuning $\Delta_{1}/\omega _{b}$ on these five bipartite entanglements  while keeping second cavity detuning $\Delta_{2}/\omega _{b}$ fixed in Fig. 4. It can be seen that for $\Delta_{2}= \omega _{b}$ i.e. when second cavity detuning is resonant with blue sideband regime, the quantity $E_{c_{1}-c_{2}}^{N}$ becomes maximum for $\Delta_{1}$ varying in the range of $0- -0.5\omega _{b}$ whereas photon hopping factor varies upto  $0-5$ although after this range  $E_{c_{1}-b_{2}}^{N}=E_{c_{2}-b_{1}}^{N}$ get finite value for both positive and negative  $\Delta_{1}/\omega _{b}$ with varying $\Gamma\kappa _{c}$ as shown in Fig. 4(a). Similarly both the quantities $E^{N}_{c_{1}-m_{2}}$ and $E^{N}_{c_{2}-m_{1}}$ get maximum for $\Delta_{1}\approx -\omega _{b}$ and after this they attain finite values again on varying $\Delta_{1}/\omega _{b}$ and $\Gamma\kappa _{c}$ as shown in Fig. 4(b) and 4(c). Moreover, the other two quantities  $E_{c_{1}-b_{2}}^{N}$ and $E_{c_{2}-b_{1}}^{N}$ attain their maximum value for $\Delta_{1}/\omega _{b}$ varying in the range of $(-1)$ to $(-2)$ even for a very high value of $\Gamma\kappa_{c}$ as shown in Fig. 4 (d) and 4(e). In another scenario for $\Delta_{2}= -\omega _{b}$ i.e. when second cavity detuning is resonant with red sideband regime, the quantity $E_{c_{1}-c_{2}}^{N}$ becomes maximum nearby to $\Delta_{1}/\omega _{b} = 0$ and  $\Gamma\approx\kappa_{c}$ however after this it decreases very rapidly on gradually increasing $\Delta_{1}/\omega _{b}$ as well as $\Gamma/\kappa_{c}$ as shown in Fig. 4(f). For this value of second cavity detuning, it can be seen that both the bipartite entanglements $E^{N}_{c_{1}-m_{2}}$ as well as $E^{N}_{c_{2}-m_{1}}$ get maximum only around $\Delta_{1}/\omega _{b} = 0$ to $\pm 1.0$  and $\Gamma/\kappa_{c}$ value lies  in between 7-10 and afterwards  both these entanglements vanish as shown in Figs. 4(g) and 4(h). However in this range of $\Delta_{1}/\omega _{b}$,  $E_{c_{1}-b_{2}}^{N}$ and $E_{c_{2}-b_{1}}^{N}$ first become maximum for single photon hopping factor  $\Gamma/\kappa_{c}$ values which are in between the range 5-7 and then both the bipartite entanglements  become zero although  a further increase in $\Gamma/\kappa_{c}$ give maximum values of $E_{c_{1}-b_{2}}^{N}$ and $E_{c_{2}-b_{1}}^{N}$ as depicted in Figs. 4(i) and 4(j).\\
\begin{figure}[hbt]
\begin{center}
\includegraphics[width=1\columnwidth,height=5in]{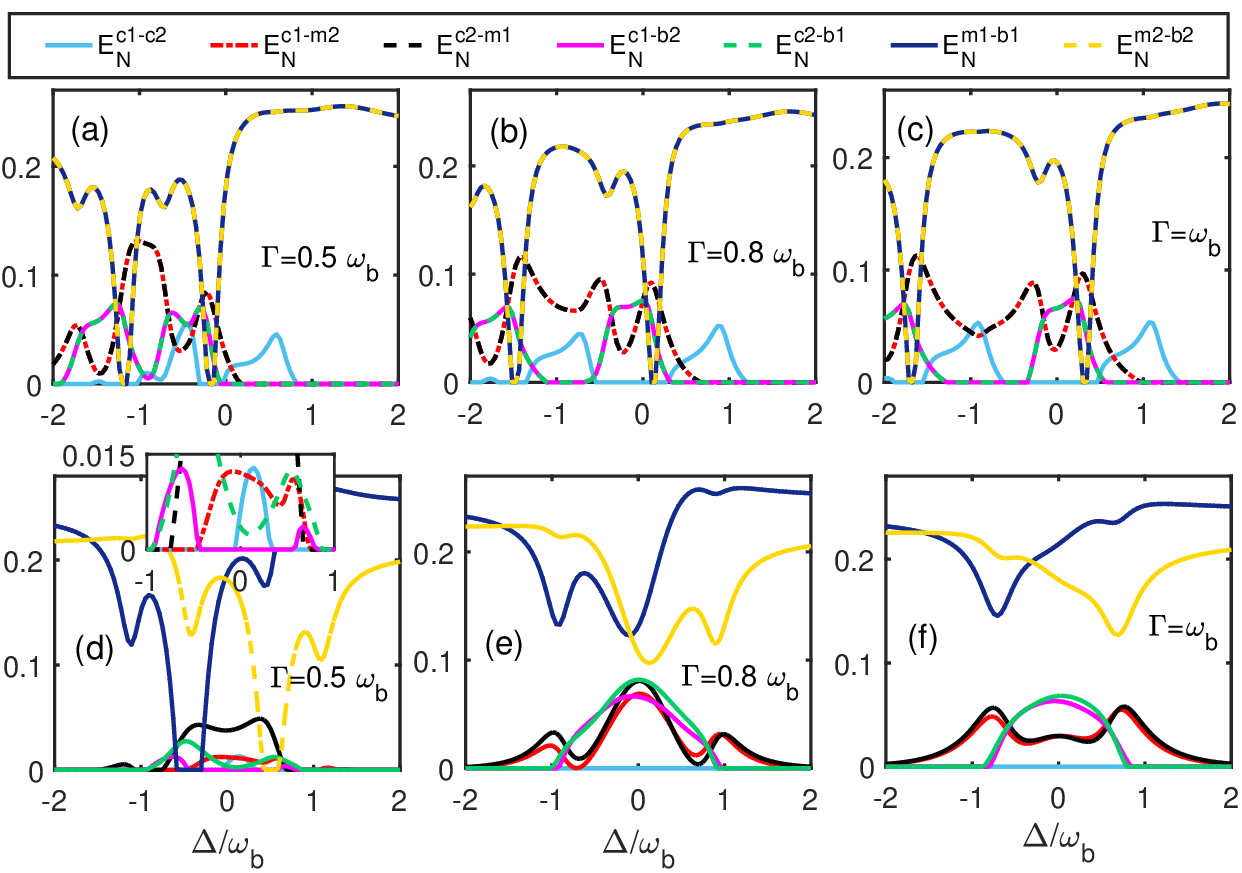}
\end{center}
\caption{(Color Online) Plot of different bipartite entanglement versus $\Delta/\protect\omega_{b}$ by taking $\Delta=\Delta_{1}=\Delta_{2}=\protect\omega_{b}$ for the upper panel and $\Delta=\Delta_{1}=-\Delta_{2}=\omega_{b}$ for the lower panel. The other parameters are as in Fig. 3.}
\end{figure}

Furthermore, in Fig. 5 we plot different distant bipartite entanglements as a function of $\Delta/\omega _{b}$   for symmetric cavities where we take $\Delta =\Delta_{1} =\Delta_{2} = \omega _{b}$ (upper panel) and for antisymmetric cavities $\Delta =\Delta_{1} =-\Delta_{2} = \omega _{b}$ (lower panel). For $\Gamma = 0.5\omega_{b}$ and symmetric cavities, the bipartite quantity $E_{c_{1}-c_{2}}^{N}$ varies from 0 to 0.6 with gradually changing  the normalised detuning $\Delta/\omega _{b}$ in between -1 to 1 and after this range the quantity  $E_{c_{1}-c_{2}}^{N}$ becomes zero as shown in Fig. 5(a). However, the other bipartite quantities $E_{m_{2}-b_{2}}^{N}$ $(E_{m_{1}-b_{1}}^{N})$ varies from 0 to 0.2 for negative values of $\Delta/\omega _{b}$ and for $\Delta/\omega _{b}$ greater than zero both of these quantities get saturated to a finite positive value. Moreover, the other bipartite quantities $E_{c_{1}-b_{2}}^{N}$ ($E_{c_{2}-b_{1}}^{N}$) as well as $E_{c_{1}-m_{2}}^{N}$ ($E_{c_{2}-m_{1}}^{N}$) become almost zero for positive values of $\Delta/\omega _{b}$ as shown in Fig. 5(a). It can be seen that for this value of $\Gamma$ a significant amount of entanglement transfer takes place from $E_{m_{2}-b_{2}}^{N}$ $(E_{m_{1}-b_{1}}^{N})$ to $E_{c_{1}-b_{2}}^{N}$ ($E_{c_{2}-b_{1}}^{N}$) and $E_{c_{1}-m_{2}}^{N}$ ($E_{c_{2}-m_{1}}^{N}$) at  $\Delta/\omega _{b}$ $\approx$ -0.3 and -1.2. For $\Gamma = 0.8\omega_{b}$, the bipartite entanglement $E_{c_{1}-c_{2}}^{N}$ become finite for $\Delta/\omega _{b}$ varying in the range of (0.3)-(1.3) and  (-0.5)-(-1.3)  as shown in Fig. 5(b). It can be also seen that the bipartite quantities $E_{m_{2}-b_{2}}^{N}$ $(E_{m_{1}-b_{1}}^{N})$ almost get around   0.2 for positive as well as negative values of $\Delta/\omega _{b}$ except for certain values of $\Delta/\omega _{b} \approx 0.2$ and $-1.5$ whereas $E_{c_{1}-m_{2}}^{N}$ ($E_{c_{2}-m_{1}}^{N}$) has finite values upto $\Delta/\omega _{b} \approx 0.5$ and $E_{c_{1}-b_{2}}^{N}$ ($E_{c_{2}-b_{1}}^{N}$) becomes zero even for negative values of $\Delta/\omega _{b}$. In this case we get maximum entanglement transfer  from $E_{m_{2}-b_{2}}^{N}$ $(E_{m_{1}-b_{1}}^{N})$ to $E_{c_{1}-b_{2}}^{N}$ ($E_{c_{2}-b_{1}}^{N}$) and $E_{c_{1}-m_{2}}^{N}$ ($E_{c_{2}-m_{1}}^{N}$) around $\Delta/\omega _{b}$ $\approx$ 0.1 and -1.5. If we increase further single photon hopping factor upto $\Gamma = \omega_{b}$ then the quantity $E_{c_{1}-c_{2}}^{N}$ remains finite  only for  $\Delta/\omega _{b}$ varying in the range of $(0.5)-(1.5)$ as well as $(-0.5)-(-1.5)$ whereas $E_{m_{2}-b_{2}}^{N}$ $(E_{m_{1}-b_{1}}^{N})$ almost gets  around  0.25 except at certain values of the $\Delta/\omega _{b}$ for which the entanglement transfer takes place between different bipartite correlations as shown in Fig. 5(c). In this case $E_{c_{1}-m_{2}}^{N}$ ($E_{c_{2}-m_{1}}^{N}$) has finite values upto $\Delta/\omega _{b} \approx 1.0$ however $E_{c_{1}-b_{2}}^{N}$ ($E_{c_{2}-b_{1}}^{N}$) qualitatively remains the same as depicted in Fig. 5(c). Moreover, the maximum entanglement transfer  from $E_{m_{2}-b_{2}}^{N}$ $(E_{m_{1}-b_{1}}^{N})$ to $E_{c_{1}-b_{2}}^{N}$ ($E_{c_{2}-b_{1}}^{N}$) and $E_{c_{1}-m_{2}}^{N}$ ($E_{c_{2}-m_{1}}^{N}$) takes place around values $\Delta/\omega _{b}$ $\approx$ 0.5 and -1.8. Now for antisymmetric cavities  $\Delta =\Delta_{1} =-\Delta_{2} = \omega _{b}$  and $\Gamma = 0.5\omega_{b}$ it can be seen that  both  the bipartite entanglements  $E_{m_{2}-b_{2}}^{N}$ $(E_{m_{1}-b_{1}}^{N})$ have finite values with  a varying $\Delta/\omega _{b}$ although for few values both becomes zero s shown in Fig. 5(d). All other bipartite entanglements have very small values for this value of $\Gamma$. For $\Gamma = 0.8\omega_{b}$ the bipartite entanglement  $E_{c_{1}-c_{2}}^{N}$ becomes  zero whereas  the quantities $E_{m_{2}-b_{2}}^{N}$ $(E_{m_{1}-b_{1}}^{N})$ have finite values from 0.1-0.25 as shown in Fig. 5(d). Moreover, the bipartite entanglements   $E_{c_{1}-b_{2}}^{N}$ ($E_{c_{2}-b_{1}}^{N}$) increases for this value of $\Gamma$ and  become finite for a  varying $\Delta/\omega _{b}$ in between the range of (-1)-(1) whereas $E_{c_{1}-m_{2}}^{N}$ ($E_{c_{2}-m_{1}}^{N}$) also increases and varies from 0-0.07(0.08) with $\Delta/\omega _{b}$ as depicted in Fig. 5(d). With a further increment in $\Gamma$ both the bipartite entanglements $E_{c_{1}-m_{2}}^{N}$ ($E_{c_{2}-m_{1}}^{N}$) becomes finite over whole range of  varying $\Delta/\omega _{b}$ whereas all other bipartite entanglements qualitatively remains the same (like earlier case of  $\Gamma = 0.8\omega_{b}$) as shown in Fig. 5(f).
\section{Conclusions}
We present an experimentally feasible scheme based on  coupled magnomechanical system where two microwave cavities are coupled through single photon hopping parameter $\Gamma$ and  each cavity also contains a magnon mode and phonon mode. We have investigated continuous variable entanglement between distant bipartitions  for an appropriate set of both cavities and  magnons detuning and their decay rates. Hence,  it can be seen that  bipartite entanglement between indirectly coupled systems are substantial in our proposed scheme. Moreover, cavity-cavity coupling strength also plays a key role in the degree of
bipartite entanglement and its transfer among different direct and indirect modes. This scheme may prove to be significant for processing continuous variable quantum information in quantum memory protocols.

\section*{Appendix A. Stability of the system}

It is worthy to discuss the stability of the subject system, since the
stability of strongly magnomechanical system is difficult to achieve. In
this section, we are going to discuss the stability of our system. The
system becomes stable only when all the eigenvalues of the drift matrix $%
\mathcal{M}$ have negative real parts. In case the sign of the real part of
eigenvalues changes, system will becomes unstable. Hence, in order to
provide the intuitive picture for the parameter regime where stability
occurs can be obtained from the Routh-Hurwitz criterion, we have plotted the
maximum of the real parts of the eigenvalues $\lambda_{\mathcal{M}}$ ($|%
\mathcal{M}-\lambda_{\mathcal{M}}\mathbb{1}|=0$) of the drift matrix $%
\mathcal{M}$ vs the normalized detunings in Fig. 7 \cite{RHCr}. It is
important to mention here that the system becomes unstable if the maximum of
the real parts of the eigenvalues $\lambda_{\mathcal{M}}$ is greater than
zero. Figure. 7 clearly shows the maximum of the real parts of the
eigenvalues $\lambda_{\mathcal{M}}$ remains negative for the chosen
parameters and hence the system is stable. Therefore, the whole set of
numerical parameters used throughout the manuscript satisfies the stability
conditions and hence the working regime we chose is the regime of stability.


\section*{Data availability}

All data used during this study are available within the article.
\section*{References}
\bibliography{RRef} 
\end{document}